\documentclass{PoS}

\title{keV sterile neutrino Dark Matter}

\ShortTitle{keV sterile neutrino Dark Matter}

\author{\speaker{Alexander Merle}%
         \thanks{Based on work done in collaboration with Marco Castellano, Marco Drewes, Andrea Grazian, Thierry Lasserre, Manfred Lindner, Steve King, Johannes K\"onig, Nicola Menci, Susanne Mertens, Viviana Niro, Norma Sanchez, Daniel Schmidt, Aurel Schneider, Maximilian Totzauer, and Matteo Viel.}\\
        Max Planck Institute for Physics, F\"ohringer Ring 6, 80805 Munich, Germany\\
        E-mail: \email{amerle@mpp.mpg.de}}

\abstract{
We give an overview of the current status of keV sterile neutrino Dark Matter. After a short introduction, we start by a general discussion of non-thermal production of Dark Matter, which applies to the three most commonly discussed mechanisms to produce sterile neutrino Dark Matter in the Universe: non-resonant, resonant, and decay production. The main goal in each case is to compute the momentum distribution function $f(p,t)$,  which incorporates the full information about the Dark Matter setting under consideration, at least in what concerns its cosmological aspects. While some difficulties lie in the actual computation of this quantity, it is decisive to obtain bounds from cosmic structure formation, which turn out to be the most crucial ones to distinguish different types of production. We will introduce these bounds and we put the resulting limits into a proper context, thereby illustrating that a significant amount of relevant parameter space is available, part of which is testable in particular by Lyman-$\alpha$ data.
}

\FullConference{Neutrino Oscillation Workshop\\
                 4 - 11 September, 2016\\
                 Otranto (Lecce, Italy)}

\begin{document}

%%%%%%%%%%%%%%%%%%%%%%%%%%%%%%%%%%%%%%%%%%%%%
\section{\label{sec:intro}Introduction}
%%%%%%%%%%%%%%%%%%%%%%%%%%%%%%%%%%%%%%%%%%%%%

If there is one thing the astro- and/or particle physics communities agree on, it is that one of the major questions of contemporary science is about the true nature of Dark Matter (DM). This substance makes up more than 80\% of the full matter content of the Universe, but up to now we have not yet figured out what is behind. While we are somewhat certain of several aspects of DM (its energy density, its rough distribution today, and the fact that it was crucial for cosmic structure formation), there are other things we do not know precisely (its identity, its production mechanism, its momentum distribution after production). While there are strong hints towards DM being made out of new elementary particles, our partial lack of knowledge suggests that we should question old believes and possibly re-evaluate certain statements made in the literature.

In the literature, the ``standard'' candidate for DM is regarded to be a Weakly Interacting Massive Particle (WIMP). However, while WIMPs do have a lot of theory motivation behind them, we cannot ignore the fact that we have not yet seen an unambiguous signal. On the contrary, on-going experiments push our limits further and further, such that, given the lack of a detection, we must start to seriously consider alternative possibilities.

Taking a step back, we should ask which non-WIMP particle ticks all the boxes for DM. One natural candidate is a \emph{sterile neutrino} $N$. This particle with mass $m_N$ is electrically neutral and interacts even more feebly than an ``active'' neutrino $\nu_\alpha$ (with flavours $\alpha = e, \mu, \tau$), namely by vertices suppressed by the small active-sterile mixing angle $\theta_\alpha$. Given that sterile neutrinos can be rather massive, they would certainly be able to act as DM, provided that they are produced in the early Universe in the right amount and with a suitable momentum spectrum~\cite{Merle:2013gea}.

Before we start, let us remark that the most complete source of information and references in existence at the moment is the \emph{White Paper on keV sterile neutrino Dark Matter}~\cite{Adhikari:2016bei}.

%%%%%%%%%%%%%%%%%%%%%%%%%%%%%%%%%%%%%%%%%%%%%
\section{\label{sec:n-th}Non-thermal DM production and how to produce keV sterile neutrinos}
%%%%%%%%%%%%%%%%%%%%%%%%%%%%%%%%%%%%%%%%%%%%%

\begin{figure}
\hspace{-0.8cm}
\begin{tabular}{lr}
\includegraphics[width=0.485\textwidth]{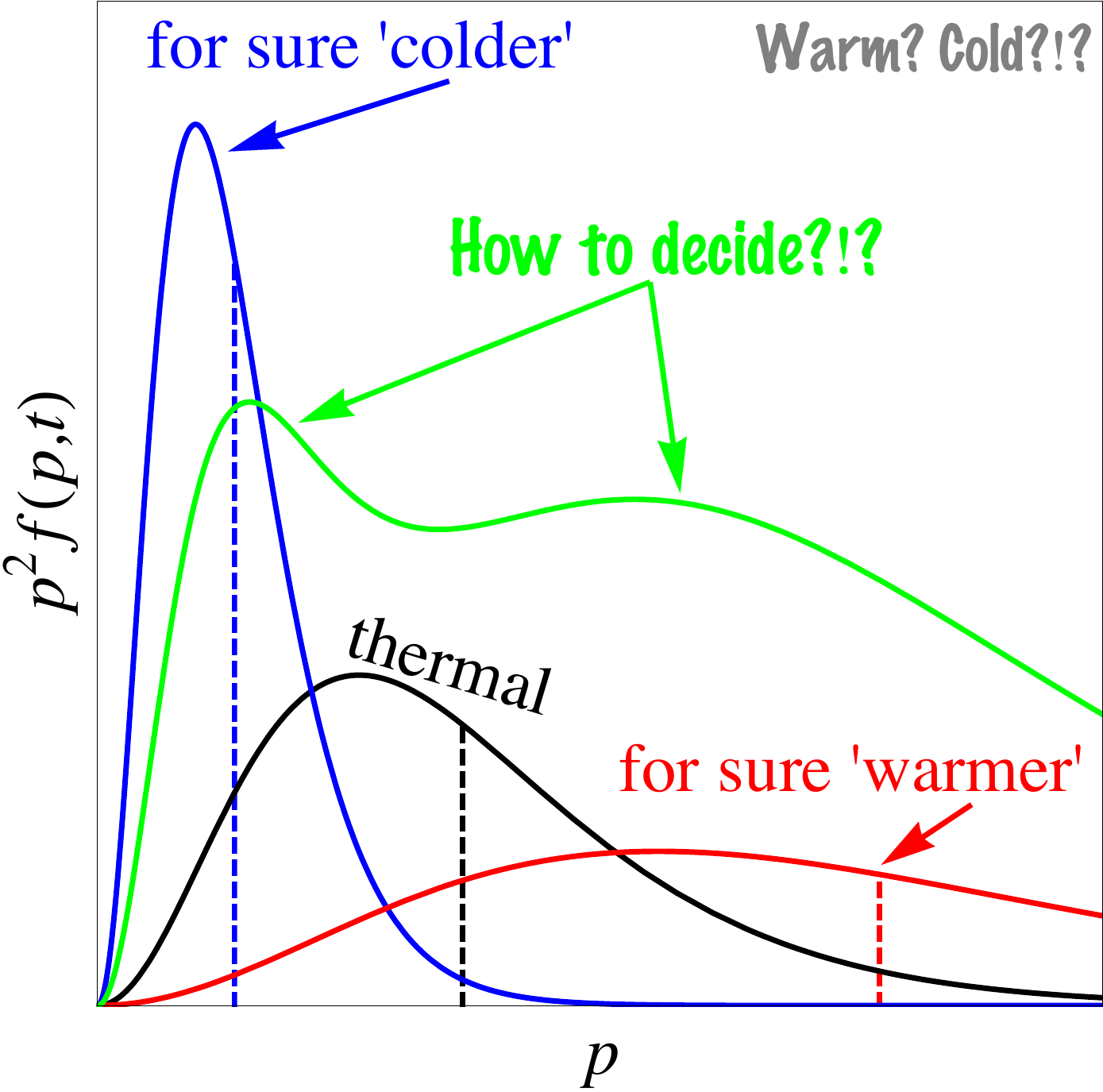} &
\includegraphics[width=0.58\textwidth]{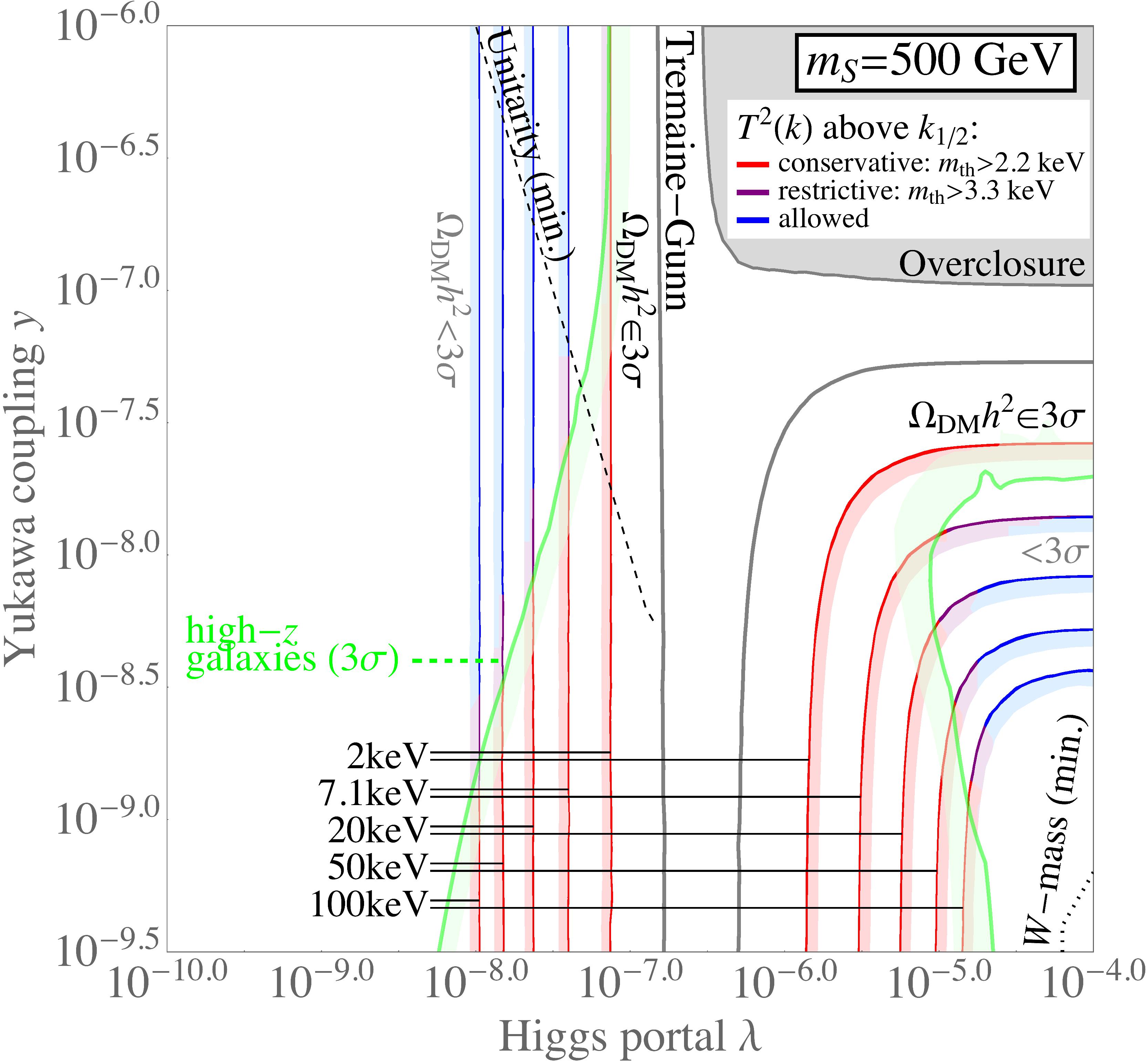}
\end{tabular}
\caption{\label{fig:dist}Schematic illustration of non-thermal distribution functions (left) and impact of structure formation constraints on an example model parameter space (right), see~\cite{Langacker:1989sv}.}
\end{figure}

Studying the literature, one finds a very common statement about sterile neutrino DM which is, however, only correct in a tiny fraction of all cases. The prejudice is that sterile neutrino DM would be \emph{warm}. The origin of this statement is in parts historic, but when looking at the details it is often just not the case. Formally, a thermal distribution would amount to the following momentum spectrum,
\begin{equation}
 {\rm thermal\ spectrum:}\ \ \ f_{\rm th}(p,T) = \frac{1}{e^{\sqrt{p^2 + m^2}/T} \pm 1}
 \left\{
 \begin{array}{l}
 {\rm +\ for\ fermions,}\\
 {\rm -\ for\ bosons,}
 \end{array}
 \right.
 \label{eq:thermal-dist}
\end{equation}
where $p$ is the DM momentum, $m$ is its mass, and $T$ is its temperature. A thermal distribution is essentially featureless and is characterised well by a single scale (e.g.\ its average momentum $\langle p \rangle$ or its temperature $T$), cf.\ black curve in the left Fig.~\ref{fig:dist}. Given that, one would speak of \emph{cold}, \emph{warm}, \emph{hot} DM if $T\ll m$, $T\approx m$, $T\gg m$, respectively.

However, in nearly all cases, sterile neutrino DM features a \emph{non-thermal} momentum spectrum. Such a spectrum can have all kinds of structures, and the only restrictions are those which force it to be physical,
\begin{equation}
 {\rm non{\mbox{-}}thermal\ spectrum:}\ \ \ f_{\rm n{\mbox{-}}th}(p) >0\ \forall p, \ \ \int {\rm d}p\ p^2 f_{\rm n{\mbox{-}}th}(p) < \infty,
 \label{eq:non-thermal-dist}
\end{equation}
where the latter requirement arises from the necessity for the DM density to be finite. Note that, contrary to the thermal distribution from Eq.~(\ref{eq:thermal-dist}), no temperature-like quantity can be defined for a non-thermal distribution. Schematically, a non-thermal distribution can look very similar to a thermal one (see, e.g., the blue and red curves in Fig~\ref{fig:dist}), or it can be vastly different and, e.g., feature several momentum scales, like the green curve in Fig~\ref{fig:dist}. In any case, a single number such as the temperature is \emph{insufficient} to characterise such a distribution. This is what makes it non-trivial to decide about whether or not a given DM momentum distribution is allowed or disfavoured by cosmic structure formation.

For the case of keV sterile neutrinos, several production mechanisms are discussed, out of which all but the very last one do exhibit non-thermal spectra:
\begin{itemize}

\item \emph{non-resonant production}: First studied by Langacker~\cite{Langacker:1989sv}, linked to DM by Dodelson and Widrow~\cite{Dodelson:1993je}. Detailed computations are of a newer date~\cite{Abazajian:2001nj,Merle:2015vzu}. This mechanism nowadays known to be excluded by observations.

\item \emph{resonant production}: First studied by Enqvist and collaborators~\cite{Enqvist:1990ek}, linked to DM by Shi and Fuller~\cite{Shi:1998km}. Detailed computations are of a newer date~\cite{Abazajian:2001nj,Ghiglieri:2015jua}. This mechanism is allowed (but constrained) by observations.

\item \emph{(scalar) decay production}: Several studies are available (see, e.g., Refs.~\cite{dec-WIMP,dec-FIMP} for different cases of parent particles). This mechanism is in good agreement with observations.

\item \emph{thermal overproduction with subsequent entropy dilution}: Discussed for keV sterile neutrinos in~\cite{dilution}. Non-trivial to get in agreement with Big Bang Nucleosynthesis.

\end{itemize}
Out of these, the second and third mechanisms are discussed most frequently, which is why we will put our main focus on those two here.

%%%%%%%%%%%%%%%%%%%%%%%%%%%%%%%%%%%%%%%%%%%%%
\section{\label{sec:asm}Production of keV sterile neutrinos by active-sterile mixing}
%%%%%%%%%%%%%%%%%%%%%%%%%%%%%%%%%%%%%%%%%%%%%

The most natural approach is to produce sterile neutrinos by their admixture to the active-neutrino sector~\cite{Langacker:1989sv,Dodelson:1993je,Abazajian:2001nj,Merle:2015vzu}, which exploits the fact that any reaction producing active neutrinos can also produce steriles -- only with a much smaller probability -- as long as some active-sterile mixing angles are non-zero. This could gradually produce enough sterile neutrinos to explain the observed amount of DM, but the resulting spectrum would feature too many particles with rather large momenta, making this setting similar to a hot DM case and thus excluded by data.

However, this could change for the case of a lepton number asymmetry being present in the early Universe, a suitable value of which can resonantly enhance the transitions from active to sterile neutrinos~\cite{Abazajian:2001nj,Enqvist:1990ek,Shi:1998km,Ghiglieri:2015jua}. While the origin of such an asymmetry has to be explained at first, its effects have been computed. Basically, if the asymmetry is large enough, it produces very many sterile neutrinos at a very specific combination of momentum and temperature. Thus, in the momentum distribution function, a sharp peak appears on top of a continuous spectrum, clearly comprising a non-thermal distribution. If this peak is located at a comparatively small momentum, i.e., a big fraction of the sterile neutrinos are not very fast, the resulting spectrum should be ``cooler'' (i.e., smaller momenta are more dominant) than the one resulting from non-resonant production.

It is non-trivial to obtain statements from cosmic structure formation about non-thermal spectra, cf.\ left Fig.~\ref{fig:dist}, so that new methods have to be developed to do so. Confronting resonant production with data, it turns out that the ``coldest'' region, which would be closest to the cold DM case, is located in a region of the parameter space that is by far excluded from the X-ray bound (i.e., from not observing the photons stemming from sterile neutrino DM decay, $N \to \nu \gamma$). All parameter combinations that are left can be investigated to see whether they are compatible with cosmic structure formation. This has been done in Ref.~\cite{Schneider:2016uqi}, using a modified version of the so-called extended Press-Schechter approach. Two methods have been applied to the spectra resulting from resonant production: \emph{halo counting} (i.e., whether a given DM distribution produces at least as many small haloes as dwarf galaxies are observed) and the \emph{Lyman-$\alpha$ method} (i.e., whether the distribution of the intergalactic medium coincides with the DM distribution expected from the spectrum under consideration). The result was that, contrary to some statements in the literature which did not take into account bounds from structure formation, resonant production is pushed already by current data. Thus, depending on how aggressively the Lyman-$\alpha$ data is interpreted, resonant production may even be completely excluded. However, it is in any case correct to say that structure formation does yield a strong constraint on this production mechanism, cf.\ Fig.~\ref{fig:Summary}.

%%%%%%%%%%%%%%%%%%%%%%%%%%%%%%%%%%%%%%%%%%%%%
\section{\label{sec:dec}Production of keV sterile neutrinos by particle decays}
%%%%%%%%%%%%%%%%%%%%%%%%%%%%%%%%%%%%%%%%%%%%%

Production by particle decays is an alternative way to generate keV sterile neutrino DM, by first producing a parent particle (e.g., a singlet scalar $S$) which then decays into sterile neutrinos (e.g., $S \to N N$). Typically, the decisive parameters are the Higgs portal coupling $\lambda$, which drives the production of the scalar from Standard Model particles, and the Yukawa coupling $y$, which drives the decay. Depending on whether the scalar itself equilibrates~\cite{dec-WIMP} or not~\cite{dec-FIMP}, different (non-thermal) spectra are possible, some resembling e.g.\ the green curve plotted in the left Fig.~\ref{fig:dist}.

The currently most advanced analysis of such cases has been presented in the last Ref.~\cite{dec-FIMP}, which also confronted the resulting cases with cosmic structure formation. A resulting snapshot of the allowed parameter space, for the case of a scalar mass of $m_S = 500$~GeV, is depicted in the right Fig.~\ref{fig:dist}. Here, the right (left) part of the plot corresponds to the region of large (small) $\lambda$ or, in other words, where the scalar freezes out (freezes in) before decaying. In this plot, the lines of correct abundance (regions of sizable but too low abundance) are indicated by the solid red/purple/blue lines (bands) for different sterile neutrino masses. In both cases, the red parts are forbidden by the Lyman-$\alpha$ data. Other bounds, such as the Tremaine-Gunn or overclosure bounds are also displayed, along with indirect collider bounds from unitarity and from the $W$-boson mass correction (the latter two of which however only play a role in the most minimal setting possible).

A detailed analysis confirms the picture given. Decay production is, in general, less constraint by (and thus in better agreement with) cosmic structure formation, as can also be seen from Fig.~\ref{fig:Summary}. This statement remains true even if, after DM production by decays is completed,  non-resonant production produces a late-time correction to the spectrum~\cite{Merle:2015vzu}.

Even more interestingly, on top of being the production mechanism with the better agreement with data, decay production can lead to very involved spectra with two or partially even three different characteristic momentum scales. This feature can possibly be used to address the (in)famous small scale problems of cosmic structure formation, i.e., the missing satellite, too-big-to-fail, and cusp-core problems. In order to derive bounds (e.g.\ from the Lyman-$\alpha$ forrest), the last Ref.~\cite{dec-FIMP} developed a completely new method based on the computation of the so-called squared transfer function $T^2(k)$, which encodes information on which structures (i.e., of which size $2\pi/k$) in space are suppressed compared to the cold DM case. This method consists of checking whether the part of the transfer function above the half-mode $T^2 = 1/2$ is allowed by the Lyman-$\alpha$ data, or not, and it turns out to be extremely robust. Indeed, the green lines shown in the right Fig.~\ref{fig:dist} arise from an alternative bound related to structure formation, namely the abundance of highly redshifted (i.e., distant) galaxies. This bound derives from completely different physics but, while it is slight less stringent, it basically tracks the boundary between the red and purple regions, and thus strongly supports the results obtained in the last Ref.~\cite{dec-FIMP}.

%%%%%%%%%%%%%%%%%%%%%%%%%%%%%%%%%%%%%%%%%%%%%
\section{\label{sec:conc}Conclusions and outlook}
%%%%%%%%%%%%%%%%%%%%%%%%%%%%%%%%%%%%%%%%%%%%%

\begin{figure}
\includegraphics[width=0.9\textwidth]{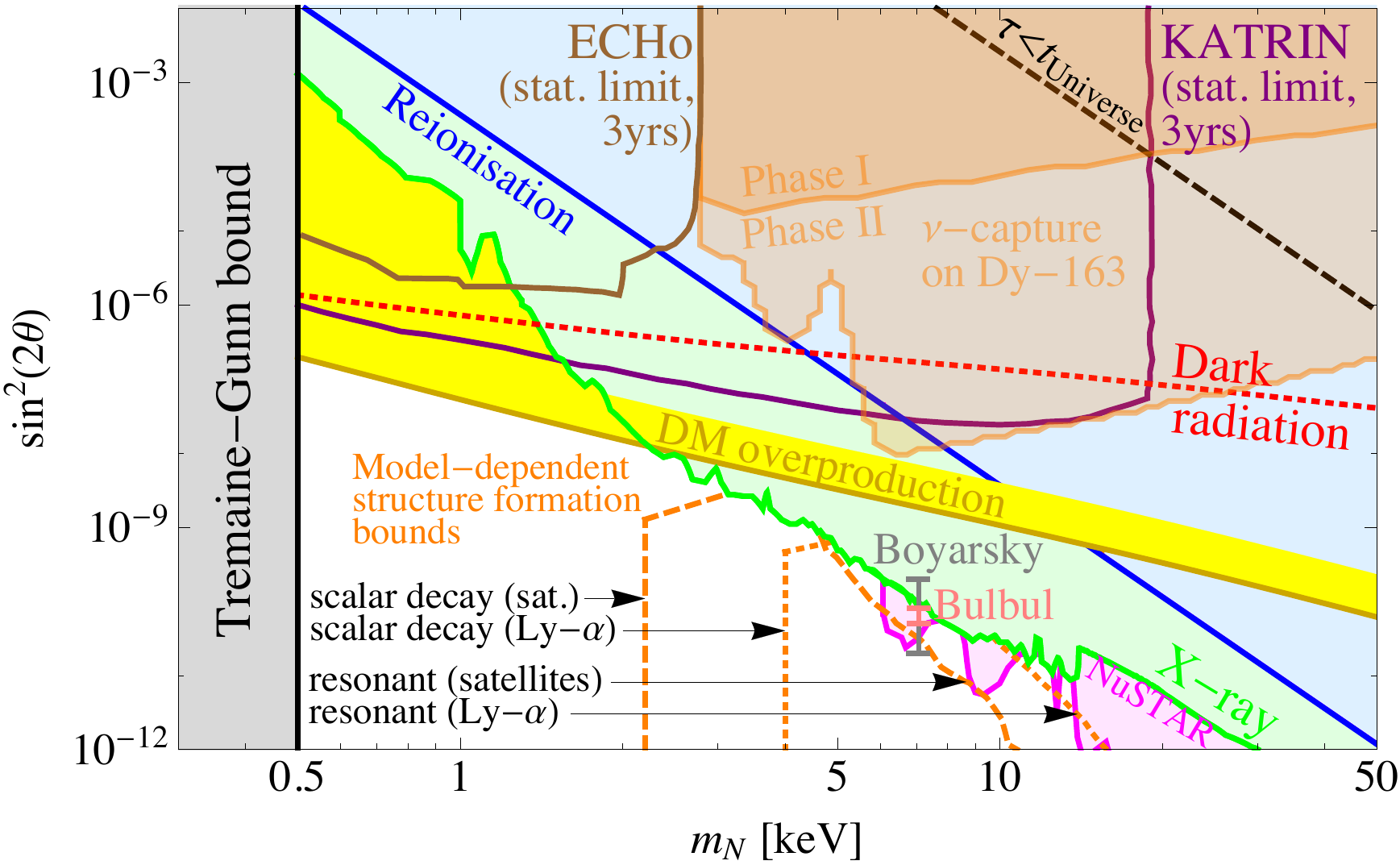}
\caption{\label{fig:Summary}Summary plot of current constraints and future experimental reaches.}
\end{figure}

We end by looking at the current situation of keV sterile neutrino DM. In Fig.~\ref{fig:Summary}, we have collected many current limits. First of all, the most stringent limits exist on the active-sterile mixing (collectively labeled $\theta$ here, because not all the limits may apply to all generations of fermions). Clearly, for large masses, the X-ray bound (green) is the strongest, due to the rate of the $N \to \nu \gamma$ decay depending on the mass $m_N$ to the fifth power. For smaller masses, though, the superior bound is the one stemming from not overproducing the DM by non-resonant production (yellow). Other bounds, such as those from reionisation (blue), dark radiation (red dotted), or the lifetime (black dashed), confirm the stronger bounds but cannot compete with them. From the other side, cosmic structure formation is the strongest driving force to constrain $m_N$. There exists a general lower bound on the sterile neutrino mass (Tremaine-Gunn bound, gray rectangle), stemming from the mere fact that sterile neutrinos are fermions, applied to the cores of galaxies. When taking into account information about the production mechanism, this bound can be strongly improved. Using the Lyman-$\alpha$ bound or the requirement of producing at least as many small haloes as we observe dwarf galaxies, we can in fact obtain very stringent constraints on the different production mechanisms. As can be seen from the plot, resonant production is in fact rather pushed by current constraints, for the reasons explained in Sec.~\ref{sec:asm}, while scalar decay production, Sec.~\ref{sec:dec}, seems in better shape. Note that, in both cases, the spectra closest to the cold DM limit have been chosen, which is a rather conservative approach.

The ways forward in this field should be clear from the plot. First and foremost, obviously, if our data on cosmic structures and, related to that, our understanding of cosmic structure formation in general improves, this may enable us to discriminate between different early Universe production mechanisms. Of course, another direction that should be pushed for is to continue hunting for the smoking gun X-ray signature stemming from DM decay. While the hopes were high for the Hitomi satellite, it was lost in 2016 through a chain of unfortunate events. Other searches such as NuSTAR (pink) can improve the limits -- however, due to many unknowns involved, one must be careful not to overstate these advances (as, e.g., done in the first version of the NuSTAR paper~\cite{Perez:2016tcq}, which incorrectly stated that it would be closing in on the entire parameter space of sterile neutrino DM, while in reality it only does so for resonant production). Finally, there are also ground-based experiments trying to constrain active-sterile mixing. While several attempts such as KATRIN/TRISTAN~\cite{Mertens:2014nha}, ECHo~\cite{ECHo}, or DyNO~\cite{Lasserre:2016eot} are on their way, they unfortunately do not seem to be able to compete with astrophysical/cosmological limits.

Nevertheless, this field clearly offers both, sufficient parameter space available and means to probe it -- plus a well-motivated non-thermal DM candidate which may in particular strongly impact on cosmic structure formation. We can be curious what the future holds for this field.

%%%%%%%%%%%%%%%%%%%%%%%%%%%%%%%%%%%%%%%%%%%%%
\section*{Acknowledgements}
%%%%%%%%%%%%%%%%%%%%%%%%%%%%%%%%%%%%%%%%%%%%%

\noindent
I would like to thank all my collaborators on keV sterile neutrinos, in particular Viviana Niro, Aurel Schneider, and Max Totzauer. Furthermore, I acknowledge partial support by the Micron Technology Foundation, Inc., and by the European Union through the Horizon~2020 research and innovation programme under the Marie Sklodowska-Curie grant agreements No.~690575 (InvisiblesPlus RISE) and No.~674896 (Elusives ITN). Of course, I also want to thank the organisers of NOW 2016 for a great meeting in an excellent environment. Finally, I guess that all participants of NOW 2016 may agree with me that we have learned a lot, in particular in what concerns tarantism.

\end{document}